\documentclass[11pt]{article}

\usepackage{amsmath,amssymb} 
\usepackage{bm}
\usepackage{graphicx} 
\usepackage{booktabs}

\usepackage[margin=1in]{geometry}

\begin{document}

\title{Identifying Central Nodes in Multiplex Networks by Embracing Layer-Specific Heterogeneity via DomiRank}

\author{
    Ru Zheng$^{1,2}$,
    Marcus Engsig$^3$,
    Alejandro Tejedor$^{2,4,5}$\thanks{Corresponding author: alej.tejedor@gmail.com},
    Yamir Moreno$^{2,4}$
}
\date{
    {\small
    $^1$ School of Mathematics and Statistics, Central South University, Changsha 410083, China \\
    $^2$ Institute for Biocomputation and Physics of Complex Systems (BIFI), Universidad de Zaragoza, Spain \\
    $^3$ Directed Energy Research Centre, Technology Innovation Institute, Abu Dhabi, UAE \\
    $^4$ Department of Theoretical Physics, University of Zaragoza, Spain \\
    $^5$ Department of Civil and Environmental Engineering, University of California, Irvine, USA \\
    }
}

\maketitle

\begin{abstract}
{The robustness and resilience of complex systems are crucial for maintaining functionality amid disruptions or intentional attacks. Many such systems can be modeled as networks, where identifying structurally central nodes is essential for assessing their  robustness and susceptibility to failure. Traditional centrality metrics often face challenges in identifying structurally important nodes in networks exhibiting heterogeneity at the network scale, with multilayer networks being a prime example of such networks. These metrics typically fail to balance the trade-off between capturing local layer-specific structures and integrating global multiplex connectivity. In this study, we extend DomiRank centrality, a metric that has been shown to effectively assess nodal importance across diverse monoplex topologies, to multiplex networks. Our approach combines layer-specific DomiRank calculations with a global contextualization step, incorporating multiplex-wide DomiRank scores to combine rankings. Through synthetic and real-world network studies, we demonstrate that our generalized DomiRank framework significantly improves the identification of key nodes in highly heterogeneous multiplex networks. This work advances centrality-based robustness assessments by addressing the fundamental trade-off between layer adaptability and multiplex-wide coherence.}

\end{abstract}

\section{\label{sec:Introduction}Introduction}
The robustness and resilience of complex systems are critical areas of study, given their importance in understanding system behavior and their practical implications across various domains \cite{Watts1998, Callaway2000,
Bashan2013extreme, Brown2006, Engsig2022, Liu2022, Artime2024}. These systems span natural ecosystems (e.g., gene-protein interactions \cite{Jeong2001,liu2020robustness}, ecological networks \cite{Dunne2002,suweis2013emergence,hervias2024structure}), social structures (e.g., human \cite{sekara2016fundamental,kumar2021evolution} and organizational interactions \cite{hidalgo2007product}), and critical infrastructure like power grids \cite{zhu2014revealing, arianos2009power, albert2004structural}, communication networks \cite{Cohen2001, Doyle2005, onnela2007structure}, and transportation systems \cite{ma2024robustness,guimera2005worldwide}. In natural and social systems, resilience ensures the ability to adapt and maintain functionality despite disruptions or dynamic changes \cite{gao2016universal,liu2022network}. Similarly, critical infrastructure systems, essential for modern societies, face risks from aging, accidents, and misuse\cite{Ahmad2024Dangers,ten2018impact,VIVEK2022urban}. Increasingly, however, these systems are also subject to deliberate attacks at both national and international levels\cite{Oberg2023,Soldi2023Monitoring}. This dual vulnerability underscores the urgent need to explore resilience mechanisms, enabling the development of strategies to protect these systems and mitigate potential disruptions effectively.

Network theory provides a unified framework for analyzing the robustness of complex systems by modeling them as networks, where nodes represent system components and links depict interactions. Robustness is typically assessed by evaluating a system's ability to maintain connectivity despite disruptions such as random failures or targeted attacks \cite{freitas2022graph}. Targeted attacks involve the sequential removal of nodes, prioritizing those of higher importance, as their removal often leads to an accelerated breakdown of system connectivity. Thus, understanding and quantifying node importance becomes essential for evaluating network robustness and designing strategies to protect or reinforce critical elements against potential failures.

Nodal importance can be evaluated using various centrality metrics \cite{Newman2018}. For example, degree centrality assesses importance locally based on the number of connections per node. Other centralities, such as eigenvector centrality and PageRank, are based on variants of random walk dynamics, while closeness and betweenness centralities are based on distances and positional relationships of nodes within the network. The effectiveness of these metrics in identifying key nodes depends on network topology \cite{liao2017ranking, Tejedor2017_AI}. For instance, degree centrality excels in hub-dominated networks, whereas betweenness and distance-based metrics perform better in spatial networks \cite{kirkley2018betweenness}.

Recently, DomiRank centrality was introduced as a versatile metric grounded in a dynamical equation interpretable through the concept of dominance \cite{engsig2024domirank}. This metric incorporates a tunable parameter $(\sigma)$, which modulates the level of competition within the dynamics, thereby balancing the influence of local (nodal-scale) and global (network-scale) information in centrality assessments. Engsig et al. \cite{engsig2024domirank} demonstrated that DomiRank centrality outperforms other metrics in identifying structurally important nodes, with targeted attacks based on DomiRank proving more effective at dismantling network structures. The flexibility of DomiRank’s parameter enables it to perform consistently well across a wide range of network types, including regular networks (e.g., lattices), random networks (e.g., Erdos-Renyi), and heterogeneous networks such as scale-free networks (e.g., Barabasi-Albert).

However, a significant challenge remains in assessing central nodes in real-world networks where heterogeneity is present across different scales. A prime example of such heterogeneity is found in multilayer networks, where different layers represent connectivity arising from distinct processes or media, resulting in inherently diverse topologies across layers \cite{kivela2014multilayer}. This structural heterogeneity complicates the assessment of central nodes, as different layers may require different centrality measures for an effective assessment \cite{boccaletti2014structure}. Furthermore, reconciling rankings derived from distinct assessments across layers poses an additional challenge.

In this paper, we address this challenge by leveraging DomiRank centrality and exploiting two of its key properties: (1) its tunable parameter, which allows it to adapt to different network topologies and extract relevant structural information, and (2) its superior performance in network dismantling, ensuring both adaptability and efficiency. Specifically, we propose a method that first computes DomiRank scores for individual layers and then ``contextualizes'' (renormalizes) these scores based on the properties of the overall DomiRank distribution across the whole multilayer network. This approach enhances the identification of key nodes in highly heterogeneous multilayer networks, improving the robustness assessment of such systems.

\section{\label{sec:Methods}Methods}

In this section, we briefly describe DomiRank centrality as recently introduced by Engsig et al. \cite{engsig2024domirank} and present the methodology proposed to adapt it to evaluate nodal importance in multiplex networks, which is generalizable to other networks with structural heterogeneities at large scale (e.g., networks exhibiting community structure where different communities display different topologies).

DomiRank centrality is a measure of nodal importance in networks based on the concept of dominance within a node's neighborhood. The dominance scores $\boldsymbol{\Gamma} = \{\Gamma_1, \Gamma_2, \dots, \Gamma_N\}$ are determined as the steady-state solution of the differential equation:  
\begin{equation}
\frac{d \boldsymbol{\Gamma}(t)}{d t}=\alpha A\left(\mathbf{1}_{N \times 1}-\boldsymbol{\Gamma}(t)\right)-\beta\boldsymbol{\Gamma}(t),
\label{DRDiffEq}
\end{equation}
where $A$ is the adjacency matrix of the network, while $\alpha$ and $\beta$ are positive real parameters controlling the dynamics of the dominance scores. This formulation can be interpreted in terms of two mechanisms: competition and relaxation. The competition term $\alpha A\left(\mathbf{1}_{N \times 1}-\boldsymbol{\Gamma}(t)\right)$ allows nodes to gain (lose) dominance by competing with neighbors that exhibit low (high) dominance scores. The competition strength is controlled by the parameter $\alpha$. On the other hand, the relaxation term $-\beta\boldsymbol{\Gamma}(t)$ ensures that dominance scores decay exponentially in the absence of competition, with the decay rate determined by the parameter $\beta$. 

The steady-state dominance scores are controlled by the parameter ratio $\sigma = \alpha / \beta$, which also determines the balance between local and global topological features. Low values of $\sigma$ emphasize local effects within neighborhoods, while higher values incorporate more global network structure \cite{engsig2024domirank}. The existence of a steady-state solution requires $\sigma$ to lie within the interval:  
\begin{equation}
\sigma \in \left(0, \frac{-1}{\lambda_N}\right),
\end{equation}
where $\lambda_N$ is the dominant negative eigenvalue of the adjacency matrix $A$.

The steady-state dominance scores have an analytical solution given by 
\begin{equation}
\boldsymbol{\Gamma} = \sigma \left(\sigma A + I_{N \times N}\right)^{-1} A \boldsymbol{1}_{N \times 1},
\label{DRAnalySol}
\end{equation}
where $I_{N \times N}$ is the identity matrix and $\boldsymbol{1}_{N \times 1}$ is a column vector of ones. However, the iterative solution provides a much more efficient computation of DomiRank centrality, with a computational complexity that scales with the number of edges, $m$ in the network, $\mathcal{O}(m)$. This property makes DomiRank suitable for analyzing large networks.  

DomiRank centrality offers a flexible framework for identifying dominant nodes in networks by tuning the parameter $\sigma$ to capture key properties at the scale of interest, from local to mesoscopic features. This adaptability enables DomiRank to be successfully applied across diverse network topologies. However, evaluating nodal importance in highly heterogeneous systems remains a significant challenge \cite{engsig2024domirank}. While the parameter $\sigma$ allows for adjustment to different topological characteristics, networks with distinct structural features in different regions may suffer from suboptimal rankings when a single global $\sigma$ value is applied. Attempting to address this issue by subdividing the network and generating localized rankings introduces the additional complexity of reconciling these rankings into a unified ordering of nodes at the global network scale. To address this limitation, we propose a novel extension of DomiRank centrality for assessing nodal importance in heterogeneous systems, such as networks with community structure, where different communities exhibit distinct topological features, or multilayer networks that encapsulate diverse substructures. This approach accounts for the internal heterogeneity of the network while producing a single, coherent ranking of nodes. 

Here, for simplicity in notation and formulation, we present our methodology for multilayer networks, focusing specifically on multiplex networks. However, the proposed framework is general and can be applied to a wide range of network types and structures without loss of generality. Recall that a multiplex network is characterized by a fixed set of nodes replicated across multiple layers, where inter-layer connections occur exclusively between corresponding node replicas \cite{kivela2014multilayer}. We utilize the supra-adjacency matrix to encode both intra-layer topologies and inter-layer couplings within a unified mathematical formalism. Thus, for a multiplex network $\mathcal{M}$ with $l$ layers and $N$ nodes per layer, the supra-adjacency matrix $\mathcal{A}$ is a block matrix given by \cite{aleta2019multilayer,artime2022multilayer}: 
\begin{equation}
\mathcal{A}=\left[\begin{array}{cccc}A^{\mathcal{L}_1} & I & \cdots & I \\ I & A^{\mathcal{L}_2} & \cdots & I \\ \vdots & \vdots & \ddots & \vdots \\ I & I & \cdots & A^{\mathcal{L}_l}\end{array}\right],
\end{equation}
\noindent where $A^{\mathcal{L}_i}$ is the adjacency matrix of the network in layer $\mathcal{L}_i$, and $I$ is the $N \times N$ identity matrix, indicating that only replica nodes are allowed to interact across layers.

With this notation in place, our proposed approach to evaluate nodal importance according to DomiRank in multiplex networks consists of a two-step estimation. We first compute individual DomiRank centralities $\boldsymbol{\Gamma}^{\mathcal{L}_i}$ for each of the layers $\mathcal{L}_i$ (communities or different substructures in the case of other topologies) of the multiplex independently, using Eq.\ref{DRDiffEq}, wherein the adjacency matrix $A^{\mathcal{L}_i}$ of the corresponding layers is utilized, and the value of $\sigma^{\mathcal{L}_i} = \frac{\alpha^{\mathcal{L}_i}}{\beta^{\mathcal{L}_i}}$ is selected according to the methodology introduced by Engsig et al. \cite{engsig2024domirank}, corresponding to the optimal value in terms of dismantling the layer's largest connected component via the ranking-based sequential attack. Additionally, we compute the DomiRank centralities $\boldsymbol{\Gamma}^{\mathcal{M}}$ for each node in the multiplex while considering the whole multiplex network $\mathcal{M}$ as a network characterized by the supra-adjacency matrix $\mathcal{A}$. The parameter $\sigma$ is selected using the same criterion applied to individual layers.

In the second step, the values of $\boldsymbol{\Gamma}^{\mathcal{L}_i}$ are normalized such that their mean values are shifted to those obtained from the calculation of $\boldsymbol{\Gamma}^{\mathcal{M}}$ over the nodes belonging to layer $\mathcal{L}_i$:
\begin{equation}
\boldsymbol{\Gamma}^{\mathcal{L}_i,\text {norm }}=\boldsymbol{\Gamma}^{\mathcal{L}_i}-\langle \boldsymbol{\Gamma}^{\mathcal{L}_i} \rangle_{\mathcal{L}_i}
+\langle\boldsymbol{\Gamma}^{\mathcal{M}}\rangle_{\mathcal{L}_i},
\label{Eq:DomiRankMultiplexNorm}
\end{equation}
where $\langle \cdot \rangle_{\mathcal{L}_i}$ denotes the mean value evaluated over the nodes in layer $\mathcal{L}_i$. Thus, all nodes in the different layers obtain a DomiRank value whose variability acknowledges the particular features of that layer, while their mean values are contextualized according to the whole multiplex structure, achieving a single final ranking encapsulating all the information. 

We will evaluate the assessment of DomiRank for different multiplex network structures using three different approaches: (i) \textit{standard} $\boldsymbol{\Gamma}^{\mathcal{M}}$: DomiRank is computed using a single value of its parameter $\sigma$, and therefore it is less sensitive to the potential heterogeneity across different layer topologies, prioritizing the overall multiplex structure; (ii) \textit{single-layer}, $\boldsymbol{\Gamma}^{\mathcal{L}_i}$: DomiRank is computed individually for each layer using a different value of its parameter $\sigma^{\mathcal{L}_i}$, adapting its assessment to the properties of each of the layers. However, this approach ignores the relative arrangement of these topologies respect to each other; (iii) \textit{multiplex normalized},  $\boldsymbol{\Gamma}^{\mathcal{L}_i,\text {norm }}$: DomiRank is evaluated  acknowledging the layer heterogeneity, using a different value of the parameter $\sigma^{\mathcal{L}_i}$ for each layer, but also contextualizing the obtained values of $\boldsymbol{\Gamma}^{\mathcal{L}_i}$ using information of the assessment of $\boldsymbol{\Gamma}^{\mathcal{M}}$ as described by Eq.\ref{Eq:DomiRankMultiplexNorm}.
The evaluation of the performance of these three versions of DomiRank is done by comparing the efficiency of targeted attack generated based on those centralities in reducing the size of the largest connected component (LCC) as a function of the attack stage. 

\section{\label{sec:Results and Discussion}Results and Discussion}
In this section, we evaluate the performance of the different versions of DomiRank introduced in Section \ref{sec:Methods}.  First, we analyze synthetic networks, focusing on exploring the role of heterogeneity on the performance of the different metrics. Next, we assess the effectiveness of DomiRank in real-world topologies, where we also compare its performance against other centrality metrics.

\subsection{\label{sec:SyntheticNets} Synthetic networks}
To evaluate the performance of DomiRank in the presence of heterogeneity at the network scale, as well as the capacity of the new methodology to improve that performance, we generate four types of two-layer multiplex networks with 1000 nodes per layer ($N=2000$) with different levels of heterogeneity. Particularly, we explore (a) a multiplex network as a control experiment with no heterogeneity introduced, consisting of two layers with Erdős-Rényi topology with the same mean degree ($\bar{k}=4$), (b) a multiplex network wherein the layers' topologies are generated with the same model (Erdős-Rényi) but with different mean degrees ($\bar{k}=4,20$), (c) a multiplex network where networks in each layer have similar mean degrees ($\bar{k}=5,6$), but very different topologies (Barabási-Albert scale-free and Random Geometric Graph), and (d) a multiplex network wherein the two layers display very different topologies in terms of structure and mean degree (Barabási-Albert scale-free with mean degree $\bar{k}=6$, and Random Geometric Graph with mean degree $\bar{k}\approx16$). For each of these multiplex networks, we evaluate the performance of the standard DomiRank ($\boldsymbol{\Gamma}^{\mathcal{M}}$), DomiRank applied to each layer individually ($\boldsymbol{\Gamma}^{\mathcal{L}_i}$), and the normalized multiplex DomiRank ($\boldsymbol{\Gamma}^{\mathcal{L}_i,\text{norm}}$) in terms of the efficiency of targeted attacks based on them in reducing the connectivity of the network. This reduction in connectivity is evaluated by the reduction of LCC of the multiplex as a function of the attack stage.

Note that to avoid biases in the evaluation and comparison of the three approaches and to ensure generalizability to multilayer networks or networks with community structure, each node of the multiplex is treated as a separate entity, and therefore, the removal of a node in a layer does not imply the removal of its replica node in the other layers. Similarly, the evaluation of the LCC is conducted by treating the whole multiplex network as a standard monoplex network.

\begin{figure*}[!h] %
    \centering
    \includegraphics[width=\textwidth]{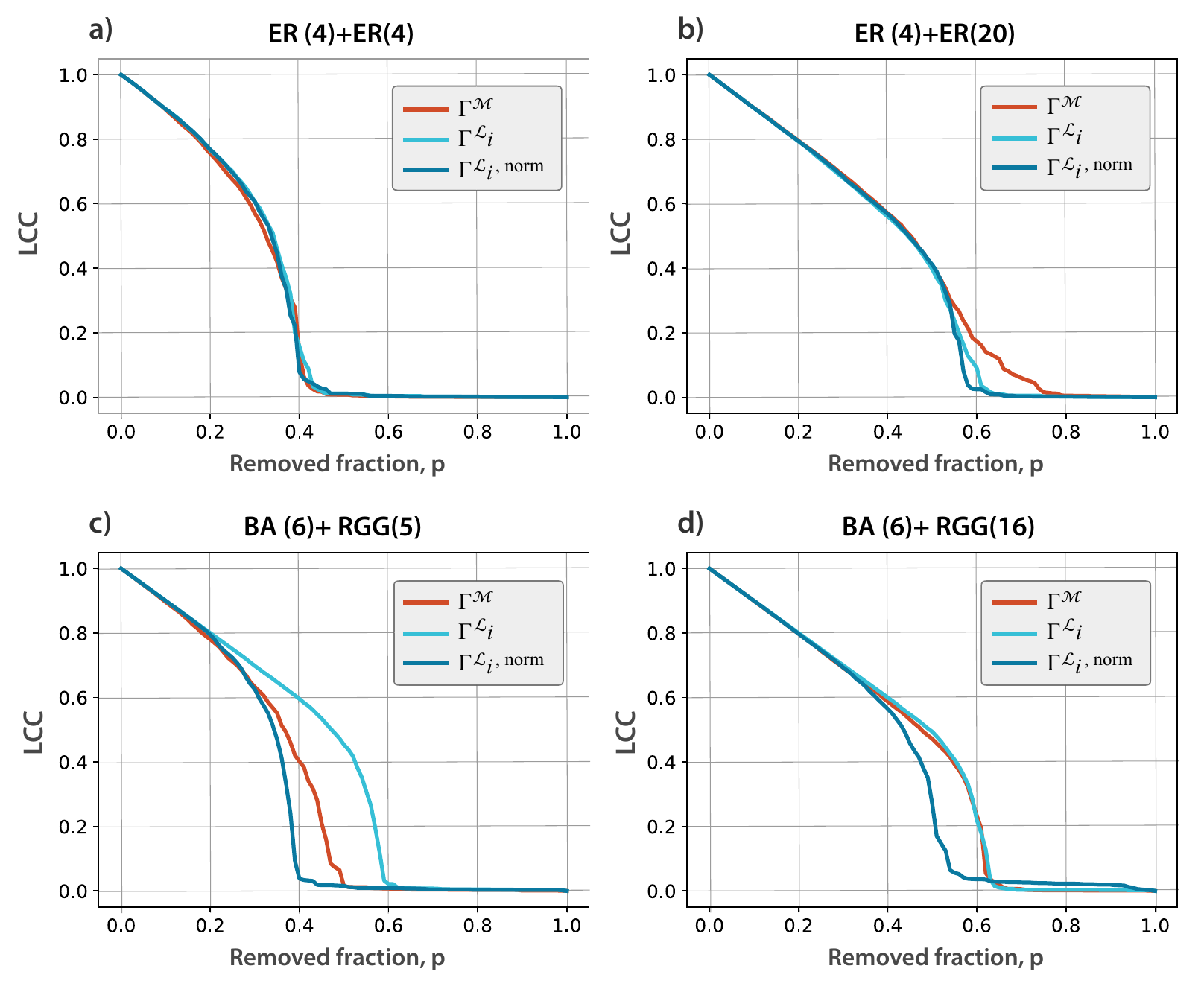} %
    \caption{Evolution of Largest Connected Component (LCC) under targeted attacks on synthetic two-layer multiplexes. Panels $\mathbf{(a)-(b)}$ illustrate the dynamics of the relative size of the LCC in two-layer Erdős–Rényi (ER) multiplex networks undergoing sequential node removal, where in panel (a), both layers exhibit the same mean degree ($\bar{k}=4$), while in panel (b), the layers exhibit degree heterogeneity ($\bar{k}=4$ and $20$). Panels $\mathbf{(c)-(d)}$ depict the evolution of the relative LCC size as a function of the remove fraction of nodes for multiplex networks composed of layers with different topologies: Barabási-Albert (BA) and random geometric graph (RGG). The mean degree of the BA topology is $\bar{k}=6$ in both panels, while the RGG has $\bar{k} \approx 5$ in panel (c) and $\bar{k} \approx 16$ in panel (d). All multiplex networks consist of 1000 nodes per layer ($N = 2000$).}
    \label{fig:SyntheticNets}
\end{figure*}

Fig. \ref{fig:SyntheticNets} shows the evolution of the LCC as a function of the attack stage for the four multiplex networks explored. A superior evaluation of nodal importance by a centrality metric should result in an attack where the LCC decreases more rapidly, leading to a smaller area under the curve (AUC). Firstly, we observe that in our control experiment (Fig. \ref{fig:SyntheticNets}a), the three strategies yield very similar results, as there is no layer heterogeneity present in the system. Despite the small differences between the three strategies, their relative ordering agrees with our expectation. Thus, the attack strategy based on $\boldsymbol{\Gamma}^{\mathcal{M}}$ is the most effective as it integrates the interlayer links in the computation of DomiRank. Conversely, the worst strategy is the one based on $\boldsymbol{\Gamma}^{\mathcal{L}_i}$, as it is completely agnostic to the effect of connectivity between layers and the relative arrangement of the topologies. A middle ground is taken by the attack strategy based on $\boldsymbol{\Gamma}^{\mathcal{L}_i,\text{norm}}$, as its normalization allows it to incorporate additional interlayer information compared to $\boldsymbol{\Gamma}^{\mathcal{L}_i}$, though it does not fully match the results from $\boldsymbol{\Gamma}^{\mathcal{M}}$. In scenarios without layer heterogeneity, $\boldsymbol{\Gamma}^{\mathcal{M}}$ effectively captures all available information through a single parameter value, $\sigma$.

On the other hand, when the multiplex system exhibits heterogeneity across the layers, attack strategies based on $\boldsymbol{\Gamma}^{\mathcal{L}_i,\text{norm}}$ are more efficient than the other two variants (see Figs. \ref{fig:SyntheticNets}b-d). It is notable that computing $\boldsymbol{\Gamma}^{\mathcal{L}_i}$ individually in each of the layers and utilizing a unique DomiRank parameter $\sigma^{\mathcal{L}_i}$ for each layer can significantly improve performance over $\boldsymbol{\Gamma}^{\mathcal{M}}$ in some configurations (e.g., see Figs. \ref{fig:SyntheticNets}b,c). However, its performance is consistently overpassed by $\boldsymbol{\Gamma}^{\mathcal{L}_i,\text{norm}}$, which leverages normalization to contextualize DomiRank values given the global multiplex structure.

\begin{figure*}[!h] %
    \centering
    \includegraphics[width=0.8\textwidth]{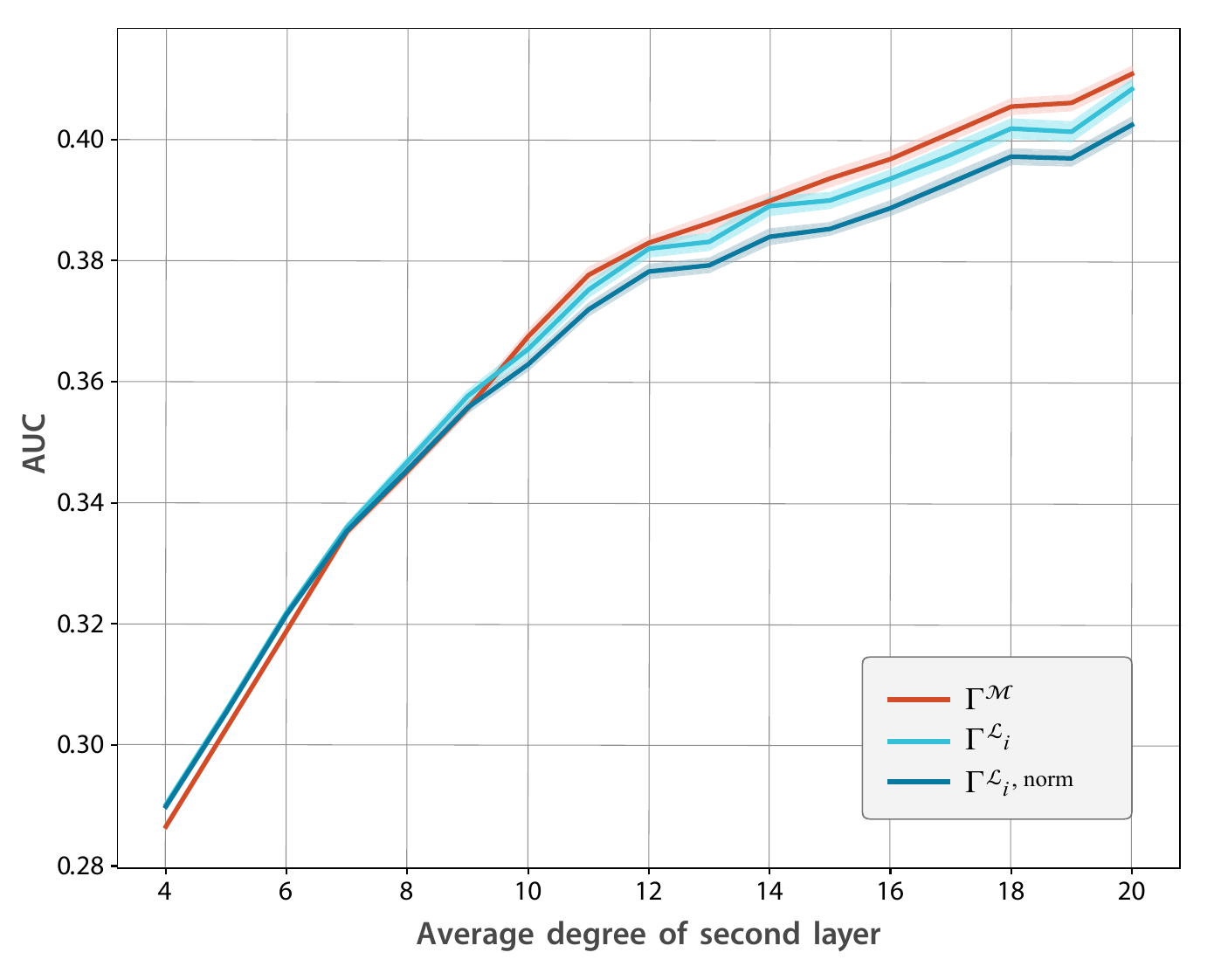} %
    \caption{Performance of the different DomiRank metrics in scenarios of increased heterogeneity. The figure shows the area under the curve (AUC) of the largest connected component evolution as a function of the degree of heterogeneity in a two-layer multiplex network with 1000 nodes per layer ($N = 2000$). Heterogeneity is controlled by varying the average degree of the Erdős–Rényi topology in the second layer, while the first layer remains fixed at $\bar{k} = 4$ (also Erdős–Rényi). Each AUC value represents the mean obtained from 100 different realizations of the multiplex network, generated with the specified model and mean degrees. The shaded regions indicate the standard error of the mean. Note that lower AUC values correspond to more effective attacks, reflecting a higher accuracy of the respective centrality metric in identifying key nodes.}
    \label{fig:explore_<k>_second_layer}
\end{figure*}

Thus, the results shown in Fig. \ref{fig:SyntheticNets} indicate that the presence and magnitude of heterogeneity across layers lead to the necessity of $\boldsymbol{\Gamma}^{\mathcal{L}_i,\text{norm}}$ to better evaluate nodal importance. To systematically assess this trade-off between global (multiplex) topological information and adaptability to layer heterogeneity, we conducted a comprehensive experiment. First, we generated two-layer Erdős-Rényi multiplex networks, keeping the mean degree of the first layer fixed at 4, while varying the average degree of the second layer from 4 to 20, i.e., gradually introducing heterogeneity. For each configuration, we evaluated the effectiveness of the different DomiRank-based attacks ($\boldsymbol{\Gamma}^{\mathcal{M}}$, $\boldsymbol{\Gamma}^{\mathcal{L}_i}$, $\boldsymbol{\Gamma}^{\mathcal{L}i,\text{norm}}$) by computing the corresponding AUC (see Fig. \ref{fig:explore_<k>_second_layer}). Note that Fig. \ref{fig:explore_<k>_second_layer} reports for each value of the average degree of the second layer, the mean AUC and its standard error (shading) for 100 different stochastic configurations of the topology. As expected, for configurations where both layers have the same degree ($\bar{k}^{\mathcal{L}_2}=4$), the three strategies yield similar results, with $\boldsymbol{\Gamma}^{\mathcal{M}}$ being slightly superior due to its ability to factor in the full multiplex structure without penalization given the lack of heterogeneity. However, as heterogeneity increases ($\bar{k}^{\mathcal{L}_2}>10$), $\boldsymbol{\Gamma}^{\mathcal{L}_i,\text{norm}}$ consistently outperforms the other approaches, effectively capturing heterogeneity while contextualizing the rankings obtained from each layer using information from the global multiplex structure.

To further demonstrate the effect of this trade-off, we present the results of a conceptually complementary experiment, wherein multiplex networks with a fixed level of heterogeneity (i.e., fixed topologies in each layer) are examined under scenarios of reduced interlayer connectivity. Particularly, we examined these system under scenarios where a fraction $p \in (0,1)$ of the interlayer links was removed. The expectation is that as $p$ increases, the potential heterogeneity in the properties of the networks in the different layers becomes more relevant in the assessment of DomiRank.

\begin{figure*}[!h] %
    \centering
    \includegraphics[width=\textwidth]{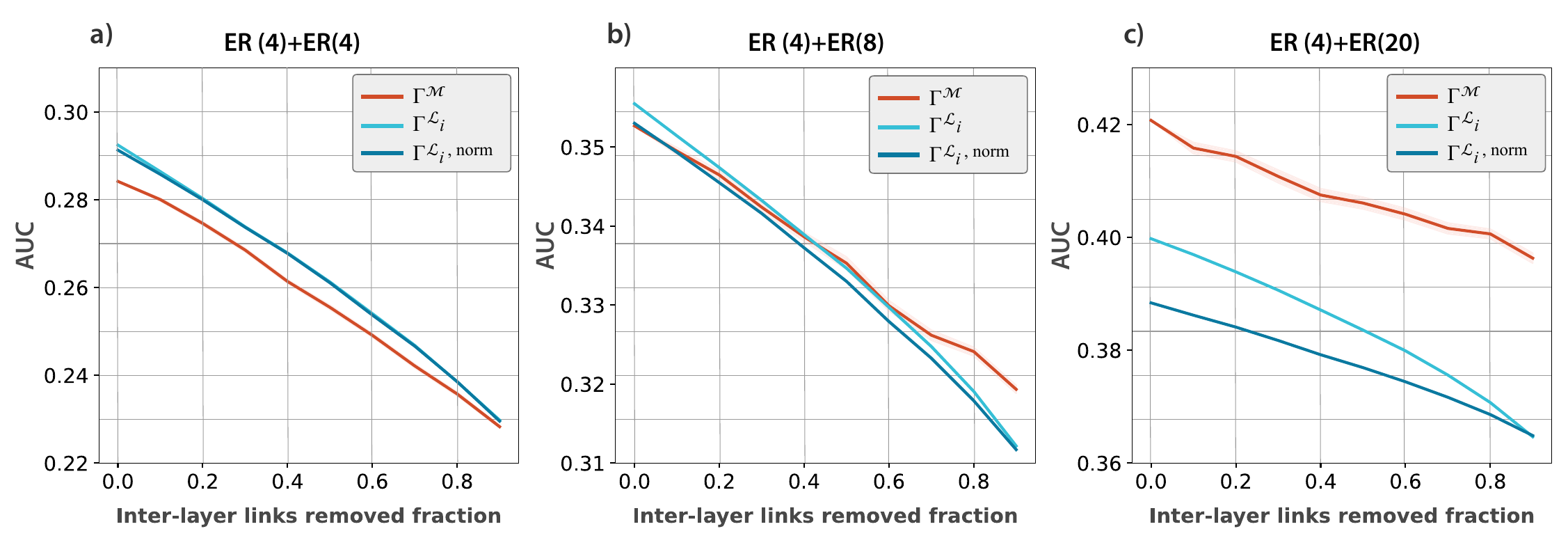} %
    \caption{Impact of inter-layer link density on DomiRank performance. The figure shows the area under the curve (AUC) of the largest connected component (LCC) evolution as a function of the fraction of inter-layer links removed in three two-layer Erdős–Rényi (ER) multiplex networks, each comprising 1000 nodes per layer. The networks differ in their layer mean degrees ($\bar{k}$): panel (a) corresponds to a multiplex network with equal mean degree in both layers ($\bar{k}=4$), panel (b) represents a multiplex with $\bar{k}=4$ and $\bar{k}=8$, and panel (c) shows results for a multiplex with layers of $\bar{k}=4$ and $\bar{k}=20$. Each AUC value represents the mean calculated from 100 different realizations, where each realization involves a randomized selection of inter-layer links removed, generated using fixed layer topologies from the specified model and mean degrees. The shaded regions indicate the standard error of the mean. Lower AUC values correspond to more effective attacks, reflecting the higher accuracy of the respective centrality metric in identifying key nodes. Notably, the strategy based on $\Gamma^{\mathcal{M}}$ exhibits a larger standard error of the mean compared to the other DomiRank variants due to its greater sensitivity to the stochasticity introduced by inter-layer connections, which influence the overall connectivity between layers.}
    \label{fig:remove inter-layer links}
\end{figure*}

The efficiency of the attacks based on the three variants of DomiRank, summarized as AUC for increasing values of $p$, are shown for three multiplex networks, wherein the first layer consists of a 1000-node Erdős-Rényi network with $\bar{k}^{\mathcal{L}_1}=4$, and the second layer also consists of a 1000-node Erdős-Rényi network with  $\bar{k}^{\mathcal{L}_2}=4$ (Fig. \ref{fig:remove inter-layer links}a), with  $\bar{k}^{\mathcal{L}_2}=8$ (Fig. \ref{fig:remove inter-layer links}b), and with  $\bar{k}^{\mathcal{L}_2}=20$ (Fig. \ref{fig:remove inter-layer links}c). 

Fig. \ref{fig:remove inter-layer links} shows the efficiency of targeted attacks based on the different versions of DomiRank as a function of the fraction $p$ of interlayer links removed.  Specifically, this figure displays the mean AUC and its standard error (shading) corresponding to the deterioration of the LCC as a result of a sequential nodal removal. Thus, smaller values of AUC correspond to more efficient attack strategies, indicating that the particular centrality metric used to generate the attack is more accurate in evaluating nodal importance.  
As expected, Fig. \ref{fig:remove inter-layer links} clearly shows a descending trend of AUC as a function of $p$ for all the multiplex networks (panels a, b and c) and for the three attacks. This trend reflects that sparser interlayer link structures are more fragile in terms of maintaining the integrity of the LCC, in other words, these structures are more susceptible to fragmentation. The results shown in Fig. \ref{fig:remove inter-layer links}a support the earlier observation that for multiplex systems lacking heterogeneity across layers, $\Gamma^{\mathcal{M}}$ is the best approach to evaluate nodal importance. This is because considering whole structure (including interlayer links) provides a better evaluation since the interlayer heterogeneity is inherently small, and the influence of inter-layer links on the network structure far exceeds that of layer heterogeneity. It is notable that this holds true for the whole range of $p$, although the other two strategies reduce their gap in efficiency for very large $p$ values, as expected. At the opposite end of the spectrum, the results in Fig. \ref{fig:remove inter-layer links}c show that in the presence of enhanced heterogeneity across layers, $\Gamma^{\mathcal{L}_i, \text{norm}}$ excels with respect to the other methodologies in agreement with previous findings. In such scenarios, layer heterogeneity requires the use of different $\sigma$ parameters for each layer to capture their distinct characteristics. Thus, Fig. \ref{fig:remove inter-layer links}c shows that the strategy based on $\Gamma^{\mathcal{L}_i, \text{norm}}$ exhibits AUCs consistently lower than the other strategies, even for small values of $p$. As expected, for large values of $p$, the difference in the AUC produced by an attack based on $\Gamma^{\mathcal{L}_i}$ and $\Gamma^{\mathcal{L}_i, \text{norm}}$ decreases, as normalization no longer provides additional information when the connectivity between the two layers approaches zero. The intermediate case, represented in Fig. \ref{fig:remove inter-layer links}b, exhibits a richer behavior, displaying multiple intersection points between the curves. We observed that the AUC values for the attack based on  $\Gamma^{\mathcal{M}}$ were initially the smallest  (i.e., providing the most accurate assessment of nodal importance). However, as the value of $p$ increased and more interlayer links were removed, the attack strategies based on $\Gamma^{\mathcal{L}_i, \text{norm}}$ resulted in lower AUC values, indicating more effective attacks. The removal of additional interlayer links further amplified the difference between these two strategies, highlighting the importance of the trade-off between accounting for heterogeneity in layer topology and recognizing the interconnectivity of the layers when assessing central nodes in the overall multiplex.

\subsection{\label{sec:RealWorld}Performance on real-world networks}
To evaluate the generalizability of the performance of $\Gamma^{\mathcal{L}_i, \text{norm}}$ to more realistic topologies, we further tested it on two real-world multiplex network structures: the C. elegans connectome multiplex network \cite{chen2006wiring,de2015muxviz} and the London transport  multiplex network \cite{de2014navigability}, which represent two distinct and representative types of multiplex network structures.
The London transportation multiplex consists of three layers where the number of nodes varies considerably across layers. Additionally, it is interesting to note that there is a limited number of interlayer links (for more detailed information about the network, see Appendix A). In contrast, the C. elegans network features two layers with a comparable number of nodes and a relatively large number of interlayer links (see Appendix A for a detailed description).

\begin{figure*}[h!] %
    \centering
    \includegraphics[width=\textwidth]{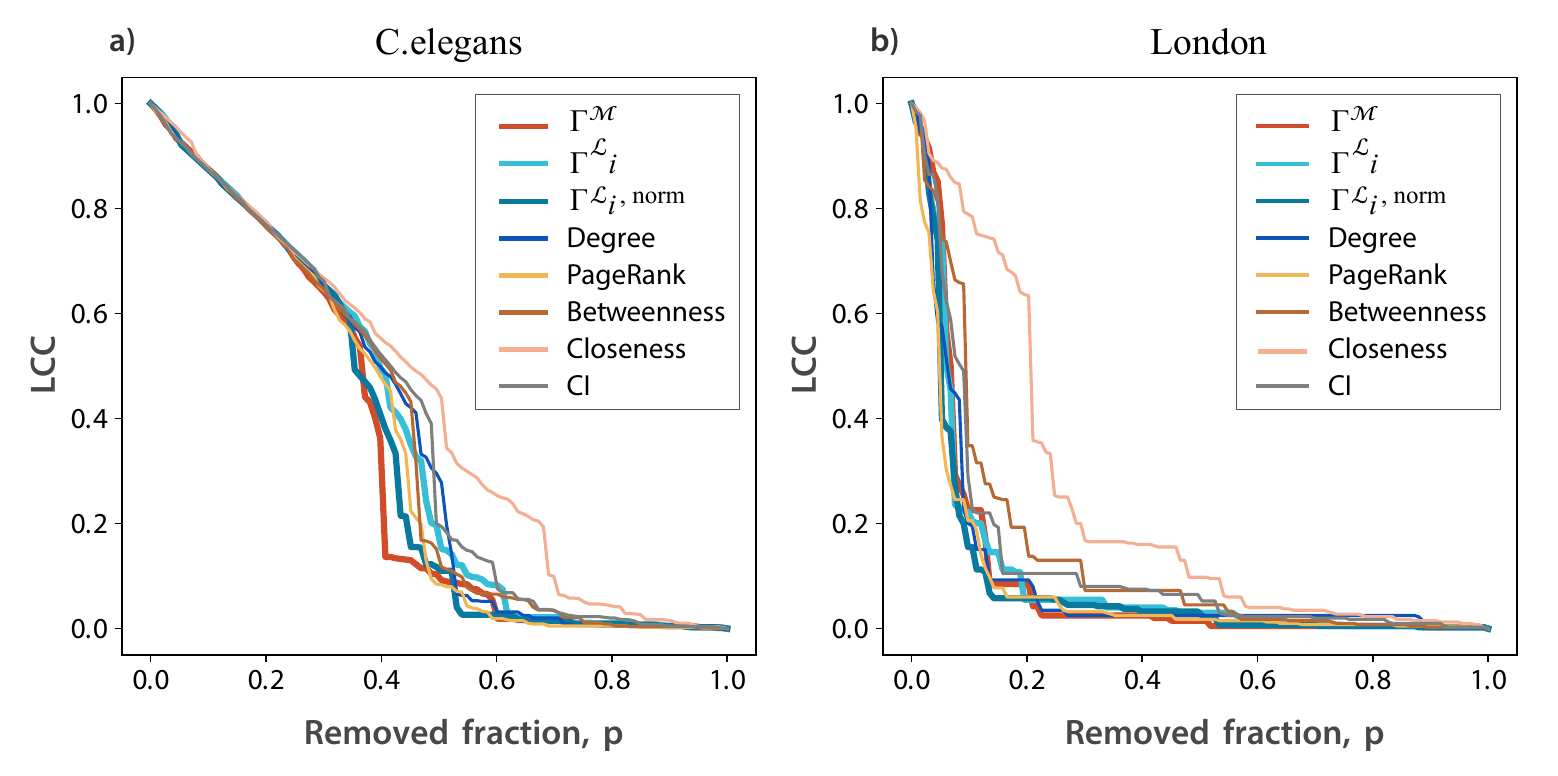} %
    \caption{Evolution of the Largest Connected Component (LCC) under targeted attacks on real-world multiplex networks. The figure shows the evolution of the LCC during sequential node removal, where nodes are removed in descending order based on different centrality metrics. Panel (a) corresponds to the C. elegans connectome multiplex network, while panel (b) represents the London multiplex transport network.}
    \label{fig:compare with baselines}
\end{figure*}

For those two real-world multiplex, we assessed the performance of the three DomiRank approaches tested in the previous sections, together with five other baseline centralities traditionally used in the literature, to contextualize their performance within a broader context. As shown in Fig. \ref{fig:compare with baselines}a, the attack based on $\Gamma^{\mathcal{M}}$ is the most effective at deteriorating the connectivity of the network. These results align with the expectations derived from the analysis of synthetic multiplexes, as C. elegans is a non-heterogeneous multiplex network with few interlayer links. On the other hand, for the London network (see Fig. \ref{fig:compare with baselines}b), the AUC of the attack generated by $\Gamma^{\mathcal{L}_i, \text{norm}}$ is consistently lower than that of the other two versions of DomiRank. These results are again consistent with the insight gained from the analysis of synthetic multiplexes, as the London multiplex network is highly heterogeneous, and therefore evaluating nodal importance benefits from calculating different $\sigma$ values for each layer.
It is interesting to notice that the $\Gamma^{\mathcal{L}_i, \text{norm}}$-based attack performs better than the attacks based on the other centralities, with the exception of PageRank which offers very similar results. Despite this similarity in terms of LCC deterioration, DomiRank has been argued \cite{engsig2024domirank} to be superior for identifying important nodes, as attacks based on DomiRank produce more intrinsic damage to the network structure. To test whether this claim also applies to multiplex networks and the new version of DomiRank, we adopted a methodology similar to that proposed by \cite{engsig2024domirank}. Specifically, we implemented an attack-recovery mechanism consisting of two steps: first, a node is removed based on its centrality-based ranking; second, a removed node is restored with probability $r$. Note, the order of node restoration is chosen to follow the original sequence of removal.

\begin{figure*}[h!] %
    \centering
    \includegraphics[width=\textwidth]{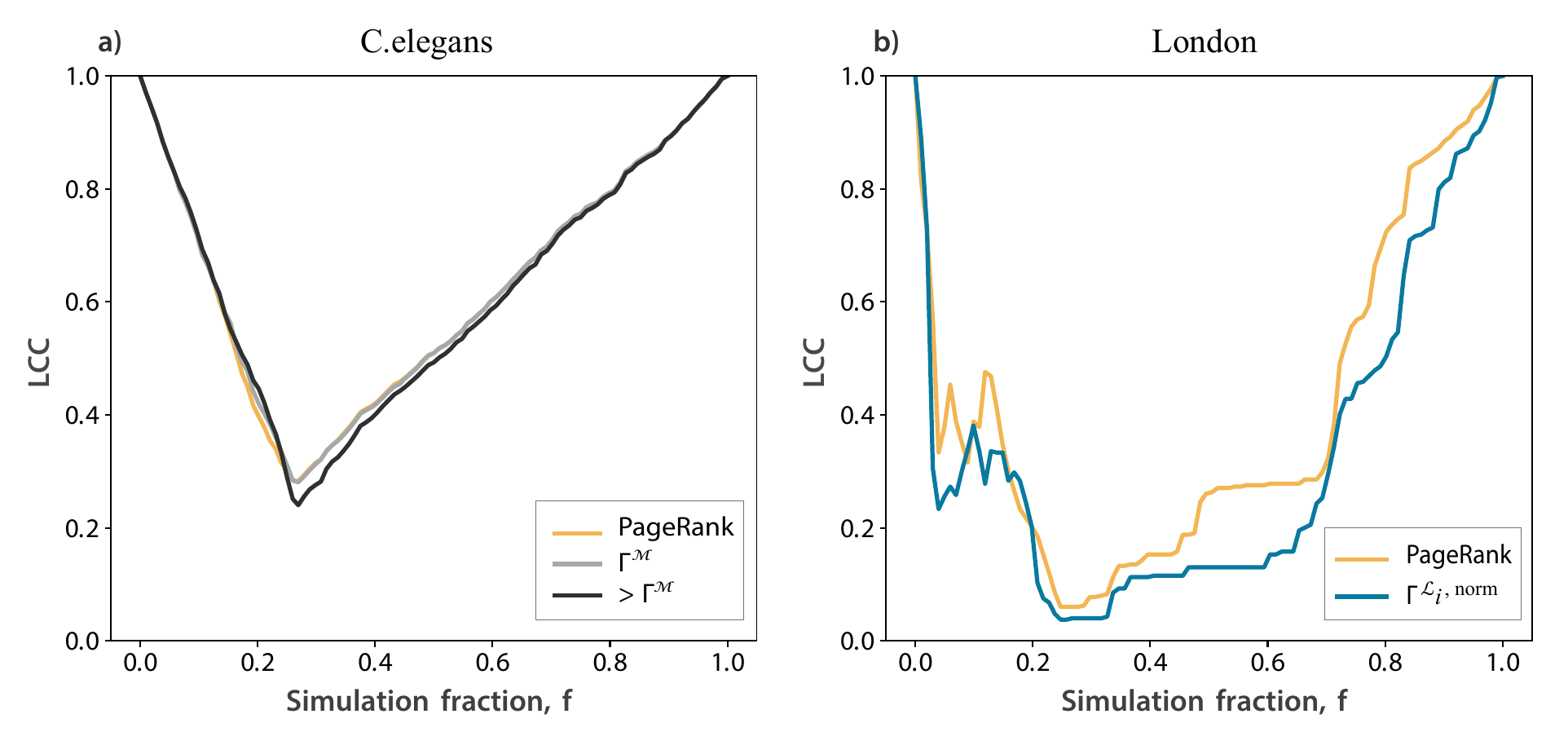} %
    \caption{Comparison of DomiRank and PageRank centralities in terms of network resilience. The effectiveness of centrality-based attacks is evaluated in a scenario incorporating a recovery mechanism for (a) the C. elegans connectome multiplex network and (b) the London multiplex transport network. The curves represent the evolution of the largest connected component (LCC) during sequential node removal, where a stochastic first-in-first-out recovery process is applied, with each removed node having a probability $r=0.25$ of being restored at each time step. Only the best-performing strategies are represented: PageRank, $\Gamma^{\mathcal{M}}$ computed for the optimal $\sigma$, and $\Gamma^{\mathcal{M}}$ under an enhanced competition scenario ($\sigma \approx 1$ - denoted in teh legend as $>\Gamma^{\mathcal{M}}$ ) in panel (a), while panel (b) compares PageRank with $\Gamma^{\mathcal{L}_i, \text{norm}}$.}
    \label{fig:Attack_recovery}
\end{figure*}

Fig. \ref{fig:Attack_recovery}a shows the evolution of the relative size of the LCC for the C. elegans multiplex network under an attack-recovery mechanism informed by three different centralities: PageRank, DomiRank $\Gamma^{\mathcal{M}}$ using a single value of the optimal parameter $\sigma$ overall C. elegans multiplex, and DomiRank $\Gamma^{\mathcal{M}}$ with enhanced dominance, wherein large values of $\sigma$ are used. The attacks based on PageRank and $\Gamma^{\mathcal{M}}$ exhibit similar performance in dismantling the LCC, but  the attack based on $\Gamma^{\mathcal{M}}$, particularly with high $\sigma$, inflicts more severe damage, as demonstrated by a slower recovery of the LCC, demosntrating the superior performance versus PageRank. An even more clear pattern emerges from a similar analysis of the the London transport multiplex. Fig. \ref{fig:Attack_recovery}b illustrates how the relative size of the LCC evolves on the London network under the attack-recovery mechanism. The results show that the attack based on $\Gamma^{\mathcal{L}_i, \text{norm}}$ is more effective and has longer-lasting effects than PageRank (and the other centrality metrics — see Appendix B), requiring a larger fraction of nodes to be recovered in the simulation to restore an equivalent LCC size compared to the PageRank strategy.

Thus, the results of the exploration of both synthetic and real-world multiplex topologies demonstrate the superior performance of DomiRank-based centralities in identifying the set of most central nodes in the system.

\section{\label{sec:conclusion}Conclusion}

In this work, we introduced and evaluated multiple variants of the DomiRank centrality metric in the context of multiplex networks, with a focus on addressing the challenges posed by the trade-off layer heterogeneity and interlayer connectivity. Through our synthetic and real-world network experiments, we demonstrated that the performance of centrality metrics is highly sensitive to the structural properties of the network and the level of heterogeneity between layers.

Our results indicate that the global multiplex DomiRank, $\boldsymbol{\Gamma}^{\mathcal{M}}$, excels in scenarios where layers share similar topological properties and interlayer connectivity is strong, as it effectively integrates information from the entire multiplex structure. However, when networks exhibit significant heterogeneity across layers, the normalized version of DomiRank, $\boldsymbol{\Gamma}^{\mathcal{L}_i, \text{norm}}$, proves to be more efficient in evaluating nodal importance. This strategy leverages normalization to incorporate local layer properties while maintaining a global perspective, providing a more robust measure of centrality when layer heterogeneity is present.

Overall, our findings highlight the key role that heterogeneity plays in our capacity to assess nodal centrality. As a centrality metric with tunable parameter, DomiRank is able to tailor the assessment of nodal importance to the dominant features of different topologies. By incorporating normalization, we offer a more flexible and effective approach for evaluating node importance in heterogeneous multiplex networks. These insights not only advance the understanding of centrality in multiplex structures but also in other network structures such as those exhibiting heterogeneous community structures. This wide applicability to topological structures that underlay many real-world complex systems, such as transportation \cite{mu2023vulnerability}, social systems \cite{murase2014multilayer}, and biological networks \cite{azevedo2021multilayer,zhao2008uncovering}, paves the way to improve our understanding of these systems not only in terms of their robustness but also in terms of the impact of heterogeneity in the dynamics that they support.

\section*{Acknowledgments}

R.Z acknowledges the support by the the Hunan Postgraduate Innovation Program (No.20220257), the China Scholarship Council Program (No.202306370262) . A.T. thanks the Spanish Ministry of Universities and the European Union Next Generation EU/PRTR for their support through the Maria Zambrano program. A.T. was also partially supported by NSF Grants EAR-2342937, RISE-2425932 and RISE-2425748. Y.M and A.T. were partially supported by the Government of Aragón, Spain, and ``ERDF A way of making Europe'' through grant E36-23R (FENOL), and by Ministerio de Ciencia e Innovación, Agencia Española de Investigación (MCIN/AEI/ 10.13039/501100011033) Grant No. PID2023-149409NB-I00.

\appendix
\section*{Appendix}
\setcounter{section}{0}
\renewcommand{\thesection}{\Alph{section}}

\section{Dataset Description}
In the manuscript, we analyze two distinct real-world multiplex: the London transport multiplex network, modeling urban transportation, and the C. elegans connectome network, representing a biological neural network. By comparing and contrasting these networks, we aim to assess the generalizability of improved centralities across diverse real-world network structures. 

Characterizing the properties of these networks is crucial for understanding the behavior of our proposed method on real-world networks with varying structural features. In table \ref{tab:networkfeature}, we provide detailed descriptions of each network's construction and key features, including the number of nodes ($N$), average degree ($\bar{k}$), average shortest path ($\bar{l}$), size of the largest connected component ($LCC$) and the number of connected components ($NCC$).

\begin{table}[ht]
\centering
\begin{tabular}{lcccccc}
\toprule
Dataset & $\mathcal{L}_i$ & $N$ & $\bar{k}$ & $LCC$ & $\bar{l}$ & $NCC$  \\
\midrule
          & 1 & 253 & 4.087 & 248 & 4.523 & 3 \\
C.elegans & 2 & 260 & 6.831 & 258 & 3.366 & 2 \\
          & 3 & 278 & 12.252 & 278 & 2.709 & 1 \\
\midrule
       & 1 & 271 & 2.303 & 271 & 13.963 & 1 \\
London & 2 & 83 & 2 & 83 & 13.486 & 1 \\
       & 3 & 45 & 2.044 & 45 & 8.232 & 1 \\

\bottomrule
\end{tabular}
\caption{Statistical description of network properties for two real-world multiplex topologies}
\label{tab:networkfeature}
\end{table}

The C. elegans multiplex network displays higher interlayer heterogeneity than the London network, as measured by the number of nodes and average degree. The lower $\bar{l}$) values in the C. elegans multiplex network indicate efficient communication pathways and suggest a well-connected structure. In contrast, there is a substantial difference in the number of nodes across the layers in the London multiplex network. The relatively low average degree suggests weaker connectivity within each layer, which typically results in a larger $\bar{l}$).
\begin{table}[ht]
\centering
\begin{tabular}{lcccc}
\toprule
Dataset & $\mathcal{L}_i$ & 1 & 2 & 3 \\
\midrule
          & 1 & &238 & 252  \\
C.elegans & 2 &238& & 259 \\
          & 3 & 252 & 259 &  \\
\midrule
          & 1 & &24 & 5  \\
London    & 2 &24& & 2\\
          & 3 & 5 & 2 &  \\
\bottomrule
\end{tabular}
\caption{Distribution of interlayer links in two real-world multiplex topologies}
\label{tab:num_interlayer}
\end{table}

Additionally, the structure of multiplex networks is also influenced by the number of interlayer links. Table \ref{tab:num_interlayer} details the interlayer links distribution, showing the number of links connecting each layer pair in the datasets. While each layer of the C. elegans network displays a more heterogeneous structure, the network exhibits a higher density of interlayer links. Conversely, the limited number of interlayer links in the London network restricts interaction between layers.

\section{Performance of other classical metrics under attack-recovery mechanisms}

This section provides a comprehensive comparison of the performance of other four classical metrics under the attack-recovery mechanism, including those not discussed in detail in the main text. By analyzing the recovery capabilities of various models in real-world networks, we further validate the robustness of our conclusions and provide a complete overview of the experimental results.

\begin{figure*}[htbp] %
    \centering
    \includegraphics[width=\textwidth]{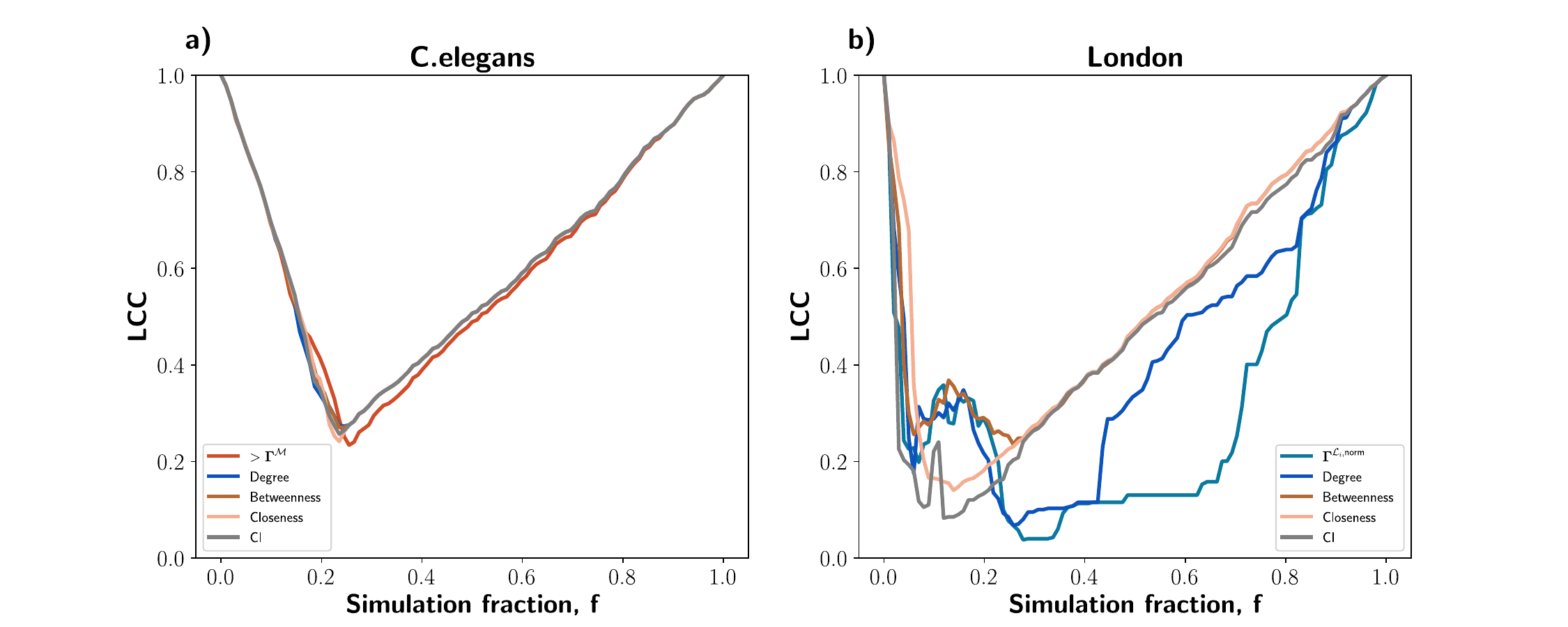} %
    \caption{Comparison of the evolution of the largest connected component (LCC) size in real-world multiplex networks under attack-recovery mechanisms based on different centrality measures.}
    \label{fig:attack_recovery_other_baselines}
\end{figure*}

The results in Fig.\ref{fig:attack_recovery_other_baselines}a illustrates the evolution of the relative size of the $LCC$ in the C.elegans multiplex network under the attack-recovery mechanisms guided by five centrality metrics: DomiRank $\Gamma^{\mathcal{M}}$ with a larger value of $\sigma$, some baselines including degree, betweenness, closeness and CI centralities. The results align with the expections derived from the analysis in Section III.B, DomiRank $\Gamma^{\mathcal{M}}$ with a larger value of $\sigma$ consistently outperforms other centralities in dismantling network strcutures.

On the other hand, for the London network, the curves in Fig.\ref{fig:attack_recovery_other_baselines}b indicate that DomiRank $\Gamma^{\mathcal{L}_i, \text{norm}}$ is superior in terms of the deterioration of the $LCC$ as a result of an attack-recovery mechanisms. These results suggest that the sequence of nodes identified by DomiRank $\Gamma^{\mathcal{L}_i, \text{norm}}$ in a highly heterogeneous real mulutiplex  network has a ongoing effect.

These findings are consistent with the observations in the main text, demonstrating the robustness of the proposed methodology and the importance of incorporating efficient recovery mechanisms in real-world multiplex networks.


\bibliographystyle{unsrt}

\begin{thebibliography}{10}

\bibitem{Watts1998}
Duncan~J. Watts and Steven~H. Strogatz.
\newblock Collective dynamics of ‘small-world’ networks.
\newblock {\em Nature}, 393:440, 1998.

\bibitem{Callaway2000}
Duncan~S. Callaway, M.~E.~J. Newman, Steven~H. Strogatz, and Duncan~J. Watts.
\newblock Network robustness and fragility: Percolation on random graphs.
\newblock {\em Phys. Rev. Lett.}, 85:5468--5471, Dec 2000.

\bibitem{Bashan2013extreme}
Amir Bashan, Yehiel Berezin, Sergey~V. Buldyrev, and Shlomo Havlin.
\newblock The extreme vulnerability of interdependent spatially embedded networks.
\newblock {\em Nature Physics}, 9(10):667--672, 2013.

\bibitem{Brown2006}
Gerald Brown, Matthew Carlyle, Javier Salmer{\'o}n, and Kevin Wood.
\newblock Defending critical infrastructure.
\newblock {\em Interfaces}, 36(6):530--544, 2006.

\bibitem{Engsig2022}
Marcus Engsig, Alejandro Tejedor, and Yamir Moreno.
\newblock Robustness assessment of complex networks using the idle network.
\newblock {\em Phys. Rev. Res.}, 4:L042050, Dec 2022.

\bibitem{Liu2022}
Xueming Liu, Daqing Li, Manqing Ma, Boleslaw~K. Szymanski, H~Eugene Stanley, and Jianxi Gao.
\newblock Network resilience.
\newblock {\em Physics Reports}, 971:1--108, 2022.
\newblock Network Resilience.

\bibitem{Artime2024}
Oriol Artime, Marco Grassia, Manlio~De Domenico, James~P. Gleeson, Hernán~A. Makse, Giuseppe Mangioni, Matjaž Perc, and Filippo Radicchi.
\newblock Robustness and resilience of complex networks.
\newblock {\em Nature Reviews Physics}, 6(2):114--131, 2024.

\bibitem{Jeong2001}
Hawoong Jeong, Sean~P Mason, A-L Barab{\'a}si, and Zoltan~N Oltvai.
\newblock Lethality and centrality in protein networks.
\newblock {\em Nature}, 411(6833):41--42, 2001.

\bibitem{liu2020robustness}
Xueming Liu, Enrico Maiorino, Arda Halu, Kimberly Glass, Rashmi~B Prasad, Joseph Loscalzo, Jianxi Gao, and Amitabh Sharma.
\newblock Robustness and lethality in multilayer biological molecular networks.
\newblock {\em Nature communications}, 11(1):6043, 2020.

\bibitem{Dunne2002}
Jennifer~A Dunne, Richard~J Williams, and Neo~D Martinez.
\newblock Network structure and biodiversity loss in food webs: robustness increases with connectance.
\newblock {\em Ecology letters}, 5(4):558--567, 2002.

\bibitem{suweis2013emergence}
Samir Suweis, Filippo Simini, Jayanth~R Banavar, and Amos Maritan.
\newblock Emergence of structural and dynamical properties of ecological mutualistic networks.
\newblock {\em Nature}, 500(7463):449--452, 2013.

\bibitem{hervias2024structure}
Sandra Herv{\'\i}as-Parejo, Mar Cuevas-Blanco, Lucas Lacasa, Anna Traveset, Isabel Donoso, Ruben Heleno, Manuel Nogales, Susana Rodr{\'\i}guez-Echeverr{\'\i}a, Carlos~J Meli{\'a}n, and Victor~M Egu{\'\i}luz.
\newblock On the structure of species-function participation in multilayer ecological networks.
\newblock {\em Nature Communications}, 15(1):8910, 2024.

\bibitem{sekara2016fundamental}
Vedran Sekara, Arkadiusz Stopczynski, and Sune Lehmann.
\newblock Fundamental structures of dynamic social networks.
\newblock {\em Proceedings of the national academy of sciences}, 113(36):9977--9982, 2016.

\bibitem{kumar2021evolution}
Aanjaneya Kumar, Sandeep Chowdhary, Valerio Capraro, and Matja{\v{z}} Perc.
\newblock Evolution of honesty in higher-order social networks.
\newblock {\em Physical Review E}, 104(5):054308, 2021.

\bibitem{hidalgo2007product}
C{\'e}sar~A Hidalgo, Bailey Klinger, A-L Barab{\'a}si, and Ricardo Hausmann.
\newblock The product space conditions the development of nations.
\newblock {\em Science}, 317(5837):482--487, 2007.

\bibitem{zhu2014revealing}
Yihai Zhu, Jun Yan, Yan Sun, and Haibo He.
\newblock Revealing cascading failure vulnerability in power grids using risk-graph.
\newblock {\em IEEE Transactions on Parallel and Distributed Systems}, 25(12):3274--3284, 2014.

\bibitem{arianos2009power}
S.~Arianos, E.~Bompard, A.~Carbone, and F.~Xue.
\newblock Power grid vulnerability: A complex network approach.
\newblock {\em Chaos: An Interdisciplinary Journal of Nonlinear Science}, 19(1):013119, 2009.

\bibitem{albert2004structural}
R{\'e}ka Albert, Istv{\'a}n Albert, and Gary~L Nakarado.
\newblock Structural vulnerability of the north american power grid.
\newblock {\em Physical review E}, 69(2):025103, 2004.

\bibitem{Cohen2001}
Reuven Cohen, Keren Erez, Daniel ben Avraham, and Shlomo Havlin.
\newblock Resilience of the internet to random breakdowns.
\newblock {\em Phys. Rev. Lett.}, 85:4626--4628, Nov 2000.

\bibitem{Doyle2005}
John~C Doyle, David~L Alderson, Lun Li, Steven Low, Matthew Roughan, Stanislav Shalunov, Reiko Tanaka, and Walter Willinger.
\newblock The “robust yet fragile” nature of the internet.
\newblock {\em Proceedings of the National Academy of Sciences}, 102(41):14497--14502, 2005.

\bibitem{onnela2007structure}
J-P Onnela, Jari Saram{\"a}ki, Jorkki Hyv{\"o}nen, Gy{\"o}rgy Szab{\'o}, David Lazer, Kimmo Kaski, J{\'a}nos Kert{\'e}sz, and A-L Barab{\'a}si.
\newblock Structure and tie strengths in mobile communication networks.
\newblock {\em Proceedings of the national academy of sciences}, 104(18):7332--7336, 2007.

\bibitem{ma2024robustness}
Junfeng Ma, Shan Ma, Xiaotian Xie, and Weihua Gui.
\newblock Robustness analysis of high-speed railway networks against cascading failures: From a multi-layer network perspective.
\newblock {\em IEEE Transactions on Network Science and Engineering}, 2024.

\bibitem{guimera2005worldwide}
Roger Guimera, Stefano Mossa, Adrian Turtschi, and LA~Nunes Amaral.
\newblock The worldwide air transportation network: Anomalous centrality, community structure, and cities' global roles.
\newblock {\em Proceedings of the National Academy of Sciences}, 102(22):7794--7799, 2005.

\bibitem{gao2016universal}
Jianxi Gao, Baruch Barzel, and Albert-L{\'a}szl{\'o} Barab{\'a}si.
\newblock Universal resilience patterns in complex networks.
\newblock {\em Nature}, 530(7590):307--312, 2016.

\bibitem{liu2022network}
Xueming Liu, Daqing Li, Manqing Ma, Boleslaw~K Szymanski, H~Eugene Stanley, and Jianxi Gao.
\newblock Network resilience.
\newblock {\em Physics Reports}, 971:1--108, 2022.

\bibitem{Ahmad2024Dangers}
Ahmad Ferdowsi, Babak Zolghadr-Asli, and Amir AghaKouchak.
\newblock Dangers of aging water infrastructure.
\newblock {\em Science}, 386(6718):158--158, 2024.

\bibitem{ten2018impact}
Chee-Wooi Ten, Koji Yamashita, Zhiyuan Yang, Athanasios~V. Vasilakos, and Andrew Ginter.
\newblock Impact assessment of hypothesized cyberattacks on interconnected bulk power systems.
\newblock {\em IEEE Transactions on Smart Grid}, 9(5):4405--4425, 2018.

\bibitem{VIVEK2022urban}
Skanda Vivek and Hannah Conner.
\newblock Urban road network vulnerability and resilience to large-scale attacks.
\newblock {\em Safety Science}, 147:105575, 2022.

\bibitem{Oberg2023}
Kolten Oberg, Ajit Srinivas, Farishta Rahman, Zongjie Wang, and Prakash Ranganathan.
\newblock Strengthening grid resilience: Lessons from the texas power blackout and implications.
\newblock In {\em 2023 North American Power Symposium (NAPS)}, pages 1--6, 2023.

\bibitem{Soldi2023Monitoring}
Giovanni Soldi, Domenico Gaglione, Simone Raponi, Nicola Forti, Enrica d'Afflisio, Paweł Kowalski, Leonardo~M. Millefiori, Dimitris Zissis, Paolo Braca, Peter Willett, Alain Maguer, Sandro Carniel, Giovanni Sembenini, and Catherine Warner.
\newblock Monitoring of critical undersea infrastructures: The nord stream and other recent case studies.
\newblock {\em IEEE Aerospace and Electronic Systems Magazine}, 38(10):4--24, 2023.

\bibitem{freitas2022graph}
Scott Freitas, Diyi Yang, Srijan Kumar, Hanghang Tong, and Duen~Horng Chau.
\newblock Graph vulnerability and robustness: A survey.
\newblock {\em IEEE Transactions on Knowledge and Data Engineering}, 35(6):5915--5934, 2022.

\bibitem{Newman2018}
Mark Newman.
\newblock {\em Networks}.
\newblock Oxford University Press, Oxford, 2 edition, 2018.

\bibitem{liao2017ranking}
Hao Liao, Manuel~Sebastian Mariani, Mat{\'u}{\v{s}} Medo, Yi-Cheng Zhang, and Ming-Yang Zhou.
\newblock Ranking in evolving complex networks.
\newblock {\em Physics Reports}, 689:1--54, 2017.

\bibitem{Tejedor2017_AI}
Alejandro Tejedor, Anthony Longjas, Ilya Zaliapin, Samuel Ambroj, and Efi Foufoula-Georgiou.
\newblock Network robustness assessed within a dual connectivity framework: joint dynamics of the active and idle networks.
\newblock {\em Scientific Reports}, 7(1):8567, Aug 2017.

\bibitem{kirkley2018betweenness}
Alec Kirkley, Hugo Barbosa, Marc Barthelemy, and Gourab Ghoshal.
\newblock From the betweenness centrality in street networks to structural invariants in random planar graphs.
\newblock {\em Nature communications}, 9(1):2501, 2018.

\bibitem{engsig2024domirank}
Marcus Engsig, Alejandro Tejedor, Yamir Moreno, Efi Foufoula-Georgiou, and Chaouki Kasmi.
\newblock Domirank centrality reveals structural fragility of complex networks via node dominance.
\newblock {\em Nature communications}, 15(1):56, 2024.

\bibitem{kivela2014multilayer}
Mikko Kivel{\"a}, Alex Arenas, Marc Barthelemy, James~P Gleeson, Yamir Moreno, and Mason~A Porter.
\newblock Multilayer networks.
\newblock {\em Journal of complex networks}, 2(3):203--271, 2014.

\bibitem{boccaletti2014structure}
Stefano Boccaletti, Ginestra Bianconi, Regino Criado, Charo~I Del~Genio, Jes{\'u}s G{\'o}mez-Gardenes, Miguel Romance, Irene Sendina-Nadal, Zhen Wang, and Massimiliano Zanin.
\newblock The structure and dynamics of multilayer networks.
\newblock {\em Physics reports}, 544(1):1--122, 2014.

\bibitem{aleta2019multilayer}
Alberto Aleta and Yamir Moreno.
\newblock Multilayer networks in a nutshell.
\newblock {\em Annual Review of Condensed Matter Physics}, 10(1):45--62, 2019.

\bibitem{artime2022multilayer}
Oriol Artime, Barbara Benigni, Giulia Bertagnolli, Valeria d'Andrea, Riccardo Gallotti, Arsham Ghavasieh, Sebastian Raimondo, and Manlio De~Domenico.
\newblock {\em Multilayer network science: from cells to societies}.
\newblock Cambridge University Press, 2022.

\bibitem{chen2006wiring}
Beth~L Chen, David~H Hall, and Dmitri~B Chklovskii.
\newblock Wiring optimization can relate neuronal structure and function.
\newblock {\em Proceedings of the National Academy of Sciences}, 103(12):4723--4728, 2006.

\bibitem{de2015muxviz}
Manlio De~Domenico, Mason~A Porter, and Alex Arenas.
\newblock Muxviz: a tool for multilayer analysis and visualization of networks.
\newblock {\em Journal of Complex Networks}, 3(2):159--176, 2015.

\bibitem{de2014navigability}
Manlio De~Domenico, Albert Sol{\'e}-Ribalta, Sergio G{\'o}mez, and Alex Arenas.
\newblock Navigability of interconnected networks under random failures.
\newblock {\em Proceedings of the National Academy of Sciences}, 111(23):8351--8356, 2014.

\bibitem{mu2023vulnerability}
Nengye Mu, Peiyuan Xin, Yuanshun Wang, Chiyao Cheng, Witold Pedrycz, and Zhen-Song Chen.
\newblock Vulnerability analysis of china’s air and high-speed rail composite express network under different node attack strategies.
\newblock {\em Annals of Operations Research}, pages 1--35, 2023.

\bibitem{murase2014multilayer}
Yohsuke Murase, J{\'a}nos T{\"o}r{\"o}k, Hang-Hyun Jo, Kimmo Kaski, and J{\'a}nos Kert{\'e}sz.
\newblock Multilayer weighted social network model.
\newblock {\em Physical Review E}, 90(5):052810, 2014.

\bibitem{azevedo2021multilayer}
Tiago Azevedo, Giovanna~Maria Dimitri, Pietro Li{\'o}, and Eric~R Gamazon.
\newblock Multilayer modelling of the human transcriptome and biological mechanisms of complex diseases and traits.
\newblock {\em NPJ systems biology and applications}, 7(1):24, 2021.

\bibitem{zhao2008uncovering}
Xing-Ming Zhao, Rui-Sheng Wang, Luonan Chen, and Kazuyuki Aihara.
\newblock Uncovering signal transduction networks from high-throughput data by integer linear programming.
\newblock {\em Nucleic acids research}, 36(9):e48--e48, 2008.

\end{thebibliography}

\end{document}